# Related Terms Search Based on WordNet / Wiktionary and its Application in Ontology Matching ♣

© Andrew Krizhanovsky

Institution of the Russian Academy of Sciences St.Petersburg Institute for Informatics and Automation RAS
andrew dot krizhanovsky@gmail.com

© Feiyu Lin

Jönköping University, Sweden
feiyu.lin@jth.hj.se

## Abstract

A set of ontology matching algorithms (for finding correspondences between concepts) is based on a thesaurus that provides the source data for the semantic distance calculations. In this wiki era, new resources may spring up and improve this kind of semantic search. In the paper a solution of this task based on Russian Wiktionary is compared to WordNet based algorithms. Metrics are estimated using the test collection, containing 353 English word pairs with a relatedness score assigned by human evaluators. The experiment shows that the proposed method is capable in principle of calculating a semantic distance between pair of words in any language presented in Russian Wiktionary. The calculation of Wiktionary based metric had required the development of the open-source Wiktionary parser software.

## 1 Introduction

Gruber [6] defined an ontology as, "*an ontology is a formal, explicit specification of a shared conceptualization*". As the main elements of the semantic web, a lot of ontologies are created in different areas and applications. Although these ontologies are developed for various purposes and domains, they always contain overlapping information. To build a collaborative semantic web, it is necessary to find ways to compare, match and integrate various ontologies. Ontology matching which is finding similar entities in the source ontologies or finding translation rules between ontologies is the first step.

There are different strategies in order to find out the similarity between entities in the current ontology matching systems. For example, these strategies can be string similarity, synonyms, structure similarity and based on instances. Synonyms strategy can help to solve the problem of using different terms in the ontologies for the same concepts. For example, an ontology may use "diagram" while the other ontology is using "graph" for the same meaning. Normally synonyms strategy is based on external resources like domain ontology, corpus, thesaurus (e.g., WordNet, Wiktionary).

Goal of this research is to compare WordNet and Wiktionary as the external data sources for a semantic distance calculation and for an ontology matching.

## 2 Ontology matching based on WordNet and Wiktionary

WordNet[1] can be treated as a partially ordered synonym resources. The total of all unique noun, verb, adjective, and adverb strings is actually 147278. WordNet consists of a set of synonyms "synsets" and "gloss" which is the definitions and examples of the concepts. A synset denotes a concept or a sense of a group of terms. Synsets provide different semantic relationships such as synonymy (similar) and antonymy (opposite), hypernymy (superconcept) / hyponymy (subconcept), meronymy (part-of), holonymy (has-a).

Semantic similarity based on WordNet has been widely explored in Natural Language Processing and Information Retrieval. These methods can be classified into three categories [13]:

• Edge-based methods: to measure the semantic similarity between two words is to measure the distance (the path linking) of the words and the position of the word in the taxonomy. For examples see Wu and Palmer [23], Resnik [18].

• Information-based statistics methods: it calculates the probability with concepts in the taxonomy first, then follows information theory. The similarity of two concepts is extent to the specific concept that subsumes them both in the taxonomy. For examples see Resnik [19], Lin [12].

• Hybrid methods: combine the above methods, e.g., X-Similarity [17], Jiang and Conrath [8], Rodriguez [20].

WordNet based semantic similarity methods can be used in two ways in the ontology matching [13]. One way is applying these methods to calculate the entities similarities in two ontologies. If two independent ontologies have a common superconcept, some methods like [20] and [17] can be used to measure structure similarity in ontology matching directly.

There are some evaluation works about WordNet based semantic similarity methods, e.g., [2] and [17].

Based on Miller and Charles [14] experiments where the results obtained for 30 pairs nouns were compared with the judgement of each pair on a scale from 0 (not similar) to 10 (total similar) by 38 students, [2] evaluates 5 methods (e.g., Resnik [19], Lin [12], Jiang and Conrath [8], etc.). [17] evaluates 14 methods (e.g., Resnik [19], Lin [12], Jiang and Conrath [8], X-Similarity [17], etc.). Both results show that Jiang and Conrath [8] method gives the best result.

At the time of writing, there are no publications on the use of Wiktionary[2] in ontology matching or related terms search. Nevertheless, one paper [24] describes application programming interfaces for Wikipedia and Wiktionary (English and German Wiktionaries).

## 3 An example of related terms search in ontology matching

To evaluate the increasing number of ontology matching methods and their qualities, OAEI (Ontology Alignment Evaluation Initiative) [3] started arranging evaluation campaigns yearly from 2004. The input of evaluation is two ontologies written in the OWL-DL language. The different elements of ontologies, e.g., concepts, instance and relations can be aligned. The usual output notations are 1:1, 1:m, n:1 or n:m. For example, one entity of one ontology can (e.g., injective, subjective and total or partial) map to entity/entities of the other ontology.

There are a lot of algorithms for semantic similarity which are used for ontology matching. There is following classification of ontology matching algorithms: internal and external [21]. Internal ontology matching algorithm exploit information which comes only with the input ontologies, external ontology matching algorithm exploit external resources such as domain ontology, corpus, thesaurus (e.g., WordNet, Wiktionary).

## 4 A measure of semantic relatedness based on the Russian Wiktionary

An experiments were conducted in order to evaluate the usefulness of the Wiktionary as a resource for related terms search, and consequently for ontology matching. It has been compared with other measures based on WordNet, Wikipedia, Roget's Thesaurus and Google.

### 4.1 Source data

The Russian Wiktionary[4] (the dump of the database as of January 2009) was parsed and the results were stored in a relational database (MySQL). So, the database of the parsed Wiktionary is a source data in the experiment. This database is compared with English and German Wiktionaries in Table 1.

The database of the parsed Wiktionary has a better coverage than WordNet (247,580 words against 150,000). At the same time, WordNet consists of over 115,000 synsets (for a total of 207,000 word-sense pairs) while the total number of semantic relations in the database of the parsed Wiktionary is about 67,000 at this moment (for a total of 177,000 word-sense pairs).

This comparison raises an interesting question: is whether the joint usage of Wiktionary and WordNet can improve the calculation of relatedness measure.

### 4.2 Evaluation based on 353 pairs of English words

WordSimilarity-353 Test Collection (353-TC) consisting of 353 pairs of English words was proposed in [4] in order to evaluate metrics and algorithms which calculates semantic similarity of words.

### 4.3 Semantic relatedness measure

The goal is to calculate a relatedness measure between two English words using Russian Wiktionary in order to estimate this measure with the help of 353-TC.

**Table 1.** The number of entries and selected types of lexical semantic information about three Wiktionaries

| | Wiktionary editions as of September 2007, from [24] | | | | A part of Wiktionary extracted by the parser. Wiktionary edition as of January 2009. | | | | |
|---|---|---|---|---|---|---|---|---|---|
| | English Wiktionary | | German Wiktionary | | Russian Wiktionary | | | | |
| | English | German | English | German | Total[5] | English | German | Russian | Ukrainian |
| Entries | 176,410 | 10,487 | 3,231 | 20,557 | 247,580 | 2,813[6] | 13,072 | 124,301 | 88,575 |
| Part of speech (POS) | | | | | | | | | |
| Nouns | 99,456 | 6,759 | 2,116 | 13,977 | 108,448 | 935 | 336 | 58,843 | 40,607 |
| Verbs | 31,164 | 1,257 | 378 | 1,872 | 26,290 | 342 | 49 | 356[7] | 24,096 |
| Adjectives | 23,041 | 1,117 | 357 | 2,261 | 26,864 | 184 | 18 | 2,168 | 23,536 |
| Unknown | POS which were not recognized by the parser | | | | 80,293 | 1,321 | 12,648 | 57,573 | 331 |
| Semantic relations | | | | | | | | | |
| Synonyms | 29,703 | 1,916 | 2,651 | 34,488 | **28,718** | 1,345 | 665 | 24,338 | 310 |
| Antonyms | 4,305 | 238 | 283 | 10,902 | 10,480 | 238 | 234 | 9,062 | 54 |
| Hypernyms | 42 | 0 | 336 | 17,286 | 18,975 | 444 | 474 | 17,033 | 115 |
| Hyponyms | 94 | 0 | 390 | 17,103 | 8,585 | 176 | 473 | 7,574 | 12 |
| Holonyms | – | – | – | – | 216 | 1 | 0 | 215 | 0 |
| Meronyms | – | – | – | – | 322 | 8 | 2 | 306 | 0 |
| Total | – | – | – | – | **67,296** | 2,212 | 1,848 | 58,528 | 491 |

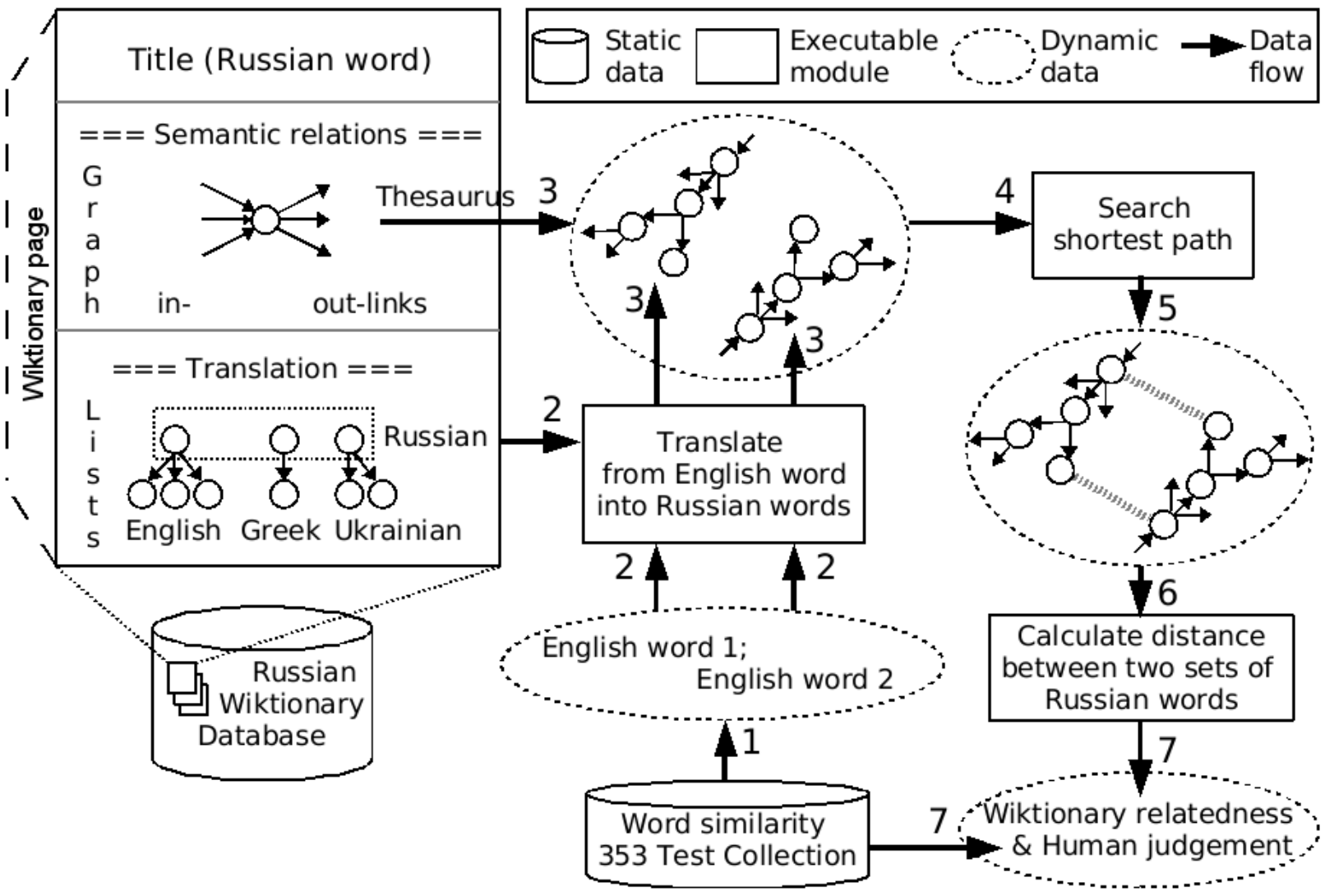

**Fig. 1.** Scheme of the experiment for calculating the semantic relatedness measure based on Russian Wiktionary data

Path based measures (Table 2) are the most attractive candidates because the Wiktionary contains thesaurus for each language (since there are Russian synonyms, English hyponyms, etc.). But the English thesaurus (in Russian Wiktionary) is much more smaller than the Russian one at this moment (2,000 against about 59,000 in Table 1). This sorrowful situation could be remedied simply by using translations from the same Wiktionary.

Thus the Russian thesaurus and translation from English to Russian (as parts of the Russian Wiktionary) will be used (see "Wiktionary page" at Fig. 1).

The process of calculation of relatedness consists of the steps illustrated in Fig. 1:

1. Take a pair of the English words from 353-TC;
2. Translate the English words to the two sets of Russian words using translations from Russian into English in the Russian Wiktionary;
3. Mark the vertices of the subgraph 1 (corresponding to the set 1 of Russian words in the thesaurus) and the subgraph 2;
4. Search the shortest paths between the marked vertices of the subgraphs 1 and 2 (Dijkstra's algorithm);
5. Extract the list of words which link two Russian words in the thesaurus, see Fig. 2 (optional step);
6. Calculate the distance between two sets of Russian words ($path^{max}_{len}$) using the paths lengths (found in step 4);
7. Save the result of the calculation for the comparison with the human judgement.

The preliminary computations include the constructing of the graph corresponding to the thesaurus and calculating the shortest paths between all pairs of nodes (for the steps 3 and 4).

Semantic relatedness measure $path^{max}_{len}$ is a maximum of lengths of shortest paths from each word of a Russian words set 1 to each word of a Russian words set 2 (Fig. 1, step 6). This measure was calculated for 353 English word pairs. The correlation with human judgements from the test collection 353-TC is 0.24 (see column "WT" in Table 2).

Let us stress that we do not make direct comparison between English words only due to the reason of a small English thesaurus in Russian Wiktionary. Notice that a development of an English Wiktionary parser will make the translation step unnecessary (for English words).

The number of cases where the distances between word pairs could not be calculated due to an absent data in the Russian Wiktionary (an absent page, or a translation, or there are no semantic relations) is 115 (32% of 353 word pairs).

### 4.4 Comparison with other metrics

The central place in the paper is occupied by Table 2 with the evaluation of semantic distance calculation algorithms and metrics against the test collection 353-TC.

The substantial part of estimations (metrics *jaccard*, *text*, $res_{hypo}$) was taken from [22]. Also $res_{hypo}$ metric was estimated in our previous paper [10]. Table 2 includes experimental data found in the following publications: [7] (the metric *jarmasz*), [4] (the search engine *IntelliZap* and *LSA* algorithm), [5] (*ESA* algorithm). The notion about the rest of the metrics could be found in other papers: the metric *wup* [23], *lch* [3] (p. 265-283), *res* [18], and *lesk* [1].

Table 2 shows correlations between the test collection 353-TC and the listed above metrics, algorithms. There are the following metrics and algorithms yielding the best results, which take into account:

I. *taxonomy structure* – 0.48, the metric *lch* [3] (English Wikipedia dataset) and 0.539, the metric *jarmasz* [7] (Roget's Thesaurus);
II. *words frequency in corpus* – 0.75, *ESA* algorithm [5] (English Wikipedia);

III. *text overlapping* – 0.21, the metric *lesk* [1] (WordNet).

Out of the scope is Green [15] algorithm (search in Wikipedia), which was not tested with 353-TC.

Table 2 shows that the Wiktionary based semantic similarity metric yields the worst result (0.24) among the path based measures (I), but it is comparable with values of text overlapping metrics (III). Best WordNet-based metrics (with value 0.34) are *lch* [3] and *res* [18].

**Table 2**. Results on correlation with human judgements of relatedness measures 353-TC to measures based on WordNet (WN), English Wikipedia (WP), Russian Wiktionary (WT)

| Dataset | **WN** | WP | **WT** | Others |
|---|---|---|---|---|
| **Metric or Algorithm** | I. Path based measures (in taxonomy) | | | |
| **wup** | 0.3 | 0.47 | – | – |
| **lch** | **0.34** | **0.48** | – | – |
| **res** $_{hypo}$ | – | 0.25-0.37[8] | – | – |
| **jarmasz** | – | – | – | **0.539** RT[9] |
| **path** $^{max}_{len}$ | – | – | **0.24** | – |
| | II. Words frequency in corpus | | | |
| **jaccard** | – | – | – | Google 0.18 |
| **res** | 0.34 | – | – | – |
| **LSA** | – | – | – | IntelliZap 0.56 |
| **ESA** | – | **0.75** | – | – |
| | III. Text overlapping | | | |
| **lesk** | 0.21 | 0.2 | – | – |
| **text** | – | 0.19 | – | – |

## 5 Implementation

The Wiktionary parser is a part of Wikokit project [10]. The software programming code is based on our previously developed Wikipedia indexing system [11] and the system that searches for related terms by analysing Wikipedia internal links [9].

The database (wikt_parsed) storing data extracted from the Wiktionary was designed using the visual tool MySQL Workbench (Fig. 3). A part of lexicographic information from Russian Wiktionary texts has been extracted and stored into this database, namely:

- a word itself (stored into the table *page*);
- a word's language and a part of speech (tables *lang_pos*, *lang*, *part_of_speech*);
- a definition (table *meaning*);
- links for key words in the definition, in the translations, in the semantic relation, i.e. in any wikified text (tables *wiki_text*, *wiki_text_words*, *page_inflection*, and *inflection*);
- semantic relations (tables *relation* and *relation_type*);
- translations (tables *translation* and *translation_entry*), where one record in the table *translation* corresponds to one meaning, and one record in the table *translation_entry* corresponds to the translation of this meaning into one language.

The developed software provides API (application programming interface) that will store and retrieve information from the database of the parsed Wiktionary. This API was used for calculating the relatedness between words in Wiktionary.

A shortest path computation on a graph (Fig. 2) was easily implemented within the Java Universal Network / Graph Framework (JUNG). The JUNG Framework is a free, open-source software library that provides a language for the manipulation, analysis, and visualization of data that can be represented as a graph or network [16].

Fig. 2 shows another problem that arises during a creation of a thesaurus from the Wiktionary data. This is a word sense disambiguation (WSD) task.

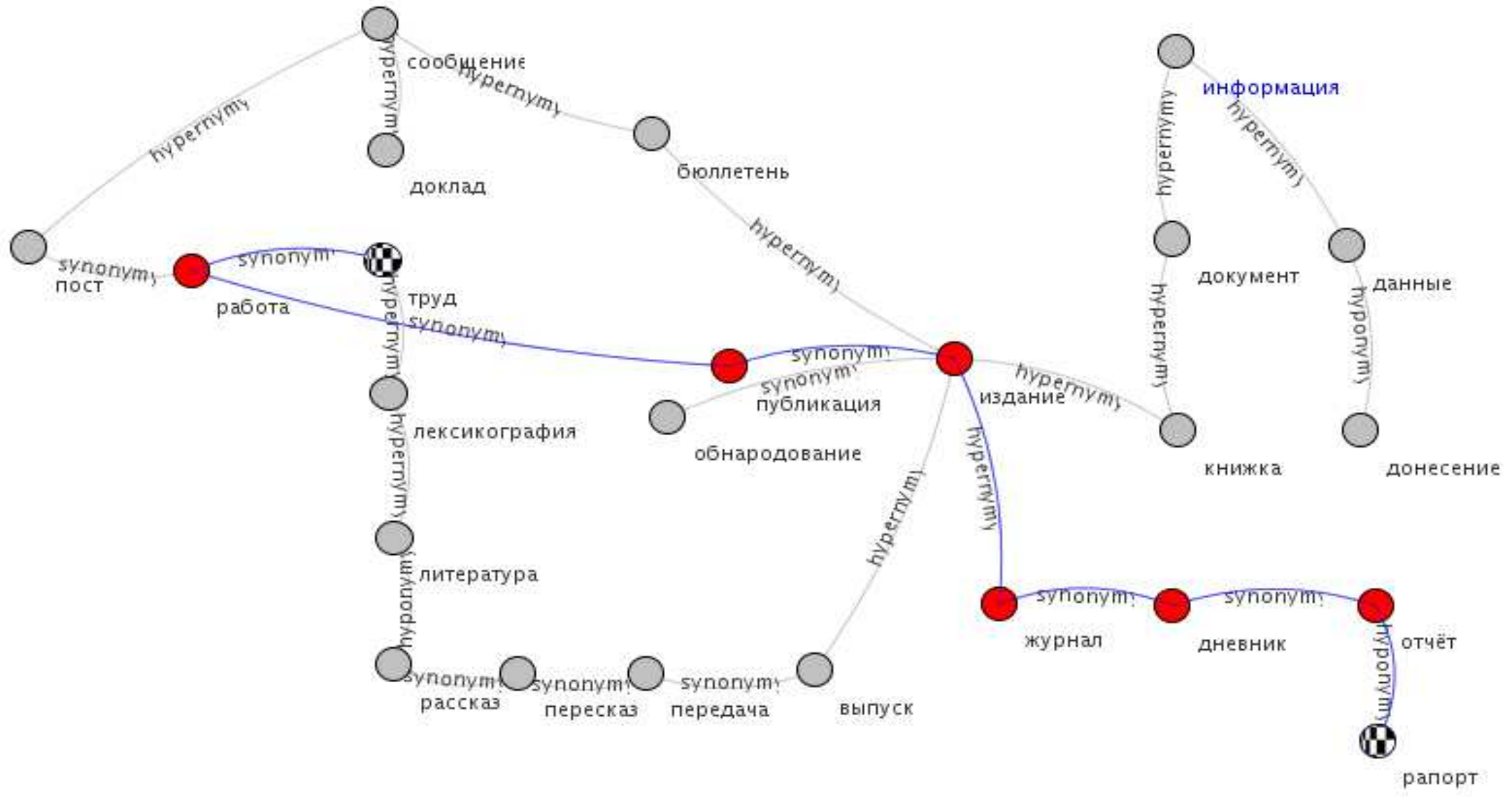

**Fig. 2.** A shortest path from a Russian word "рапорт" (raport) to a word "труд" (work, labour) found in a thesaurus of the Russian Wiktionary ("рапорт", "отчёт", "дневник", "журнал", "издание", "публикация", "работа", "труд")

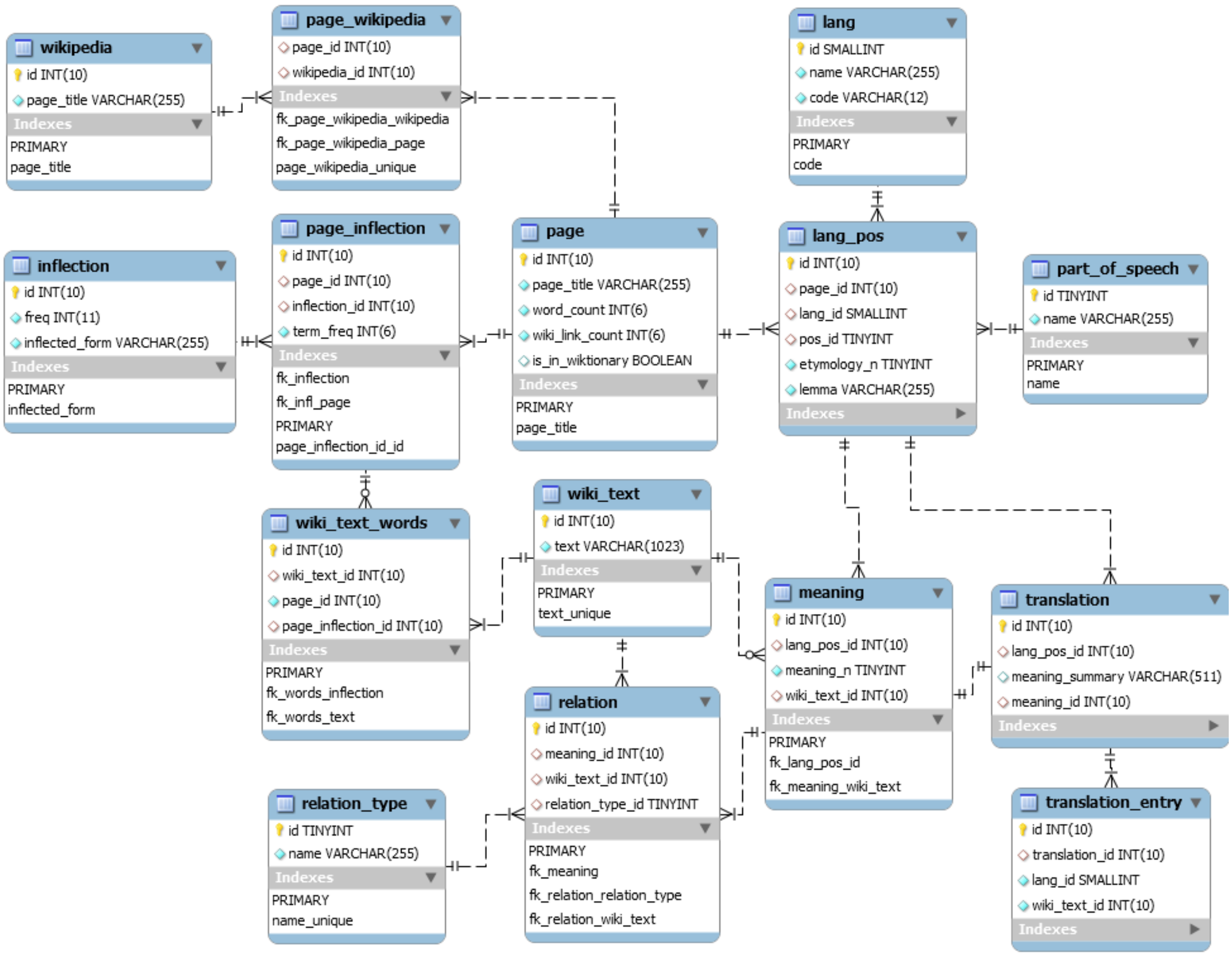

**Fig. 3.** Tables and relations in the database of the parsed Wiktionary

See the entry “журнал” in Russian Wiktionary:

1. “дневник” (journal, diary) is a near-synonym of “журнал” (journal) (in accordance with meaning number 2 of “журнал”);
2. “издание” (magazine, journal) is a hyperonym of “журнал” (in accordance with meaning number 1 of “журнал”).

The entry “дневник” (journal, diary) describes that “журнал” (journal) is a near-synonym of “дневник” with, regrettably, no number of “журнал” meaning number.

Thus, within one Wiktionary page (entry, word) the different meanings (definitions) are presented explicitly, hence the lists of synonyms, antonyms, etc. are also explicitly marked by the number of meaning. But other entries (which “mention” this word by hyperlinks in a definition, or synonym, or translation sections) do not indicate the meaning number of this entry.

It’s not a big problem for the reader, but it requires the additional worthwhile programming work of disambiguating the meanings of the words listed in the “semantic relations” section of a Wiktionary page. Thus, a WSD algorithm should be developed or adapted to the Wiktionary in order to solve this problem.

## 6 Discussion and conclusion

Wiktionary has advantage over WordNet in being of a larger size in number of words, but the database of the parsed Wiktionary has less number of relations (see Table 1). The experiment shows that a similarity calculation between two English words by the joint usage of a thesaurus and translations extracted from Russian Wiktionary could not hope to win a victory over WordNet so far.

But the present experiment shows that the proposed method is capable in principle of calculating a semantic distance between pair of words in any language presented in Wiktionary (more than 200 in Russian Wiktionary).

It should be noted that (1) other language editions of Wiktionary are out of the scope of this paper, (2) only a small part of lexicographic information from Russian Wiktionary texts has been extracted and stored into machine readable dictionary, namely:

- a word’s language,
- a part of speech,
- a definition,
- links in the definition for key words,
- semantic relations,
- and translations.

An extraction from Wiktionary of a pronunciation (phonetic transcription, a sound sample), a hyphenation, an etymology, a quotation (example sentence), a parallel text (examples with translations), a figure (which illustrates a word meaning) were not considered because this is a first step towards the creation of an open-source Wiktionary parser software.

## References

[1] S. Banerjee, T. Pedersen. An Adapted Lesk algorithm for word sense disambiguation using WordNet. In Proceedings of the Third International Conference on Intelligent Text Processing and Computational Linguistics (CICLING-02). Mexico City, February, 2002. http://www.d.umn.edu/~tpederse/Pubs/cicling2002-b.pdf

[2] A. Budanitsky, G. Hirst. Semantic distance in wordnet: An experimental, application-oriented evaluation of five measures, pp. 29–24, 2001.

[3] C. Fellbaum. WordNet: an electronic lexical database. – MIT Press, Cambridge, Massachusetts – 423 pp. – ISBN 0-262-06197-X. 1998.

[4] L. Finkelstein, E. Gabrilovich, Y. Matias, E. Rivlin, Z. Solan, G. Wolfman, E. Ruppin. Placing search in context: the concept revisited. In *ACM Transactions on Information Systems*, volume 20(1), pp. 116-131, 2002. http://www.cs.technion.ac.il/~gabr/papers/tois_context.pdf

[5] E. Gabrilovich, S. Markovitch. Computing semantic relatedness using Wikipedia-based Explicit Semantic Analysis. In Proceedings of The 20th International Joint Conference on Artificial Intelligence (IJCAI). Hyderabad, India, January, 2007. http://www.cs.technion.ac.il/~gabr/papers/ijcai-2007-sim.pdf

[6] T. R. Gruber. A translation approach to portable ontology specifications. In *Knowledge Acquisition*, volume 5(2), pp. 199-220, 1993.

[7] M. Jarmasz, S. Szpakowicz. Roget's Thesaurus and semantic similarity. In Proceedings of Conference on Recent Advances in Natural Language Processing (RANLP). Borovets, Bulgaria, pp. 212-219, 2003. http://www.nzdl.org/ELKB/

[8] J. J. Jiang and D.W. Conrath. Semantic Similarity Based on Corpus Statistics and Lexical Taxonomy. Semantic Similarity Based on Corpus Statistics and Lexical Taxonomy, 1997.

[9] A. A. Krizhanovsky. Synonym search in Wikipedia: Synarcher. In: 11-th International Conference "Speech and Computer" SPECOM'2006. Russia, St. Petersburg, pp. 474-477, 2006. http://arxiv.org/abs/cs/0606097

[10] A. A. Krizhanovsky. Evaluation experiments on related terms search in Wikipedia: Information Content and Adapted HITS (In Russian). 2007. http://arxiv.org/abs/0710.0169

[11] A. A. Krizhanovsky. Index wiki database: design and experiments. In: Corpus Linguistics, 2008. http://arxiv.org/abs/0808.1753

[12] D. Lin. An Information-Theoretic Definition of Similarity. In Proceedings of the Fifteenth International Conference on Machine Learning, 1998.

[13] F. Lin, K. Sandkuhl. A Survey of Exploiting WordNet in Ontology Matching. In Proc. IFIP AI, 2008.

[14] G. A. Miller and W. G. Charles. Contextual correlates of semantic similarity. In *Language and Cognitive Processes*, volume 6, pp. 1-28, 1991.

[15] Y. Ollivier, P. Senellart. Finding related pages using Green measures: an illustration with Wikipedia. In Association for the Advancement of Artificial Intelligence. Vancouver, Canada, 2007. http://pierre.senellart.com/publications/ollivier2006finding.pdf

[16] O'Madadhain, D. Fisher, P. Smyth, S. White, and Y.-B. Boey. Analysis and visualization of network data using JUNG (preprint). *Journal of Statistical Software*, pp. 1–35. 2007. http://jung.sourceforge.net/doc/JUNG_journal.pdf

[17] E. G. M. Petrakis, G. Varelas, A. Hliaoutakis, and P. Raftopoulou. Design and Evaluation of Semantic Similarity Measures for Concepts Stemming from the Same or Different Ontologies. In Book Design and Evaluation of Semantic Similarity Measures for Concepts Stemming from the Same or Different Ontologies, pp. 44-52, 2006.

[18] P. Resnik. Using Information Content to Evaluate Semantic Similarity in a Taxonomy. In Book Using Information Content to Evaluate Semantic Similarity in a Taxonomy, pp. 448—453, 1995.

[19] P. Resnik. Semantic Similarity in a Taxonomy: An Information-Based Measure and its Application to Problems of Ambiguity in Natural Language. In *Journal of Artificial Intelligence Research*, volume 11, pp. 95-130, 1999.

[20] M. Andrea Rodrguez and J. E. Max. Determining Semantic Similarity among Entity Classes from Different Ontologies. In *IEEE Trans. on Knowl. and Data Eng.*, volume 15(2), pp. 442-456, 2003.

[21] P. Shvaiko and J. Euzenat. A survey of schema-based matching approaches. In *Journal on Data Semantics*, (4), pp. 146–171, 2005.

[22] M. Strube, S. Ponzetto. WikiRelate! Computing semantic relatedness using Wikipedia. In Proceedings of the 21st National Conference on Artificial Intelligence (AAAI 06). Boston, Mass., July 16-20, 2006. http://www.eml-research.de/english/research/nlp/publications.php

[23] Z. Wu and M. Palmer. Verbs semantics and lexical selection. In Proceedings of the 32nd annual meeting on Association for Computational Linguistics, 1994.

[24] T. Zesch, C. Mueller, I. Gurevych. Extracting lexical semantic knowledge from Wikipedia and Wiktionary. In Proceedings of the Conference on Language Resources and Evaluation (LREC), 2008. http://elara.tk.informatik.tu-darmstadt.de/publications/2008/lrec08_camera_ready.pdf

---

♣ Part of this work was financed by the Foundation (The Swedish Institute), project CoReLib.

The research is supported partly by the project funded by grant 08-07-00264 of the Russian Foundation for Basic Research, and project 213 of the research program “Intelligent information technologies, mathematical modelling, system analysis and automation” of the Russian Academy of Sciences.

[1] WordNet, http://wordnet.princeton.edu

[2] Wiktionary, http://wiktionary.org

[3] OAEI, http://oaei.ontologymatching.org/2008

[4] Russian Wiktionary, http://ru.wiktionary.org

[5] Total, i.e. all languages in the Russian Wiktionary.

[6] Different parts of speech are considered as different entries (table *lang_pos* in the database of the parsed Wiktionary).

[7] Russian Wiktionary contains no doubt more than 356 Russian verbs, but only a part of verbs was successfully extracted by the parser.

[8] 0.25, 0.37, where 0.37 was taken from [22] (English Wikipedia, as of Feb. 2006), and 0.25 from [10] (English Wikipedia, as of May 2007).

[9] RT – Roget's Thesaurus [7].

[10] Wikokit. http://code.google.com/p/wikokit